\documentclass[11pt]{article}
\usepackage{multirow}
\usepackage{graphicx}
\usepackage{verbatim}
\usepackage{float}
\usepackage[font=footnotesize,labelfont=bf]{caption}
\usepackage[margin=1.0in]{geometry}

\title{Foil Diffuser Investigation with GEANT4}
\author{Joseph M. Fabritius II, Konstantin Borozdin, Peter Walstrom}

\begin{document}

\maketitle

\begin{abstract}
An investigation into the appropriate materials for use as a diffuser foil in electron radiography was undertaken in GEANT4. Simulations were run using various refractory materials to determine a material of appropriate Z number such that energy loss is minimal. The plotted results of angular spread and energy spread are shown. It is concluded that higher Z number materials such as tungsten, tantalum, platinum or uranium could be used as diffuser materials. Also, an investigation into the handling of bremsstrahlung, multiple coulomb scattering, and ionization in GEANT4 was performed.  
\end{abstract}

\section{Motivation}

In deciding on the best material for a diffuser foil for increasing the angular spread of the accelerator  beam in electron radiography, one must take into account four physical phenomena:
\begin{enumerate}
\item Multiple Coulomb scattering (the desired effect, which increases the angular spread of the beam).  This is mostly due to elastic scattering from nuclei and is approximately described by the Particle Data Group (PDG) formula 27.14.
\item  Ionization energy loss and straggling (the part of the energy loss not due to bremsstrahlung)
\item Energy loss due to bremsstrahlung
\item Melting temperature
\end{enumerate}

Choosing a diffuser foil for electron radiography experiments requires the material must be both refractory, that is resistant to melting or deformation under high temperatures, and to also have a low energy spread so that the effects of chromatic blur are lessened. Chromatic blurring effects can be directly seen in the energy loss of the beam through the material. By choosing a material with less energy loss an appropriate diffuser material can be found. To accomplish this task several elements were chosen from the Particle Data Group table on atomic properties of materials. Those materials that were found to have a melting point of over 1400 K were chosen for investigation: carbon (graphite), silicon, iron, tantalum, tungsten, platinum, and uranium. 

The optimal material Z was found by first choosing the foil thickness of each refractory material so that a specified angular spread, $\theta_{rms}$, was achieved. A desired angular spread of 0.2 mRad was chosen as the reference of comparison between diffuser materials. The diffuser thickness was calculated using the multiple scattering distribution equation:

\begin{equation}
\theta_{rms} = \frac{13.6 \textnormal{MeV}}{\beta c p}z\sqrt{\frac{x}{X_{0}}}\left[1+0.038\frac{x}{X_{0}}\right]
\end{equation}

\noindent where  $\beta c$ is the electron speed, $z$ is the particle charge number, $x$ is the thickness of the material, and $X_{0}$ is the radiation length of the material. The main material dependence is in the factor $\sqrt{\frac{x}{X_0}}$, where $x$ is the thickness in g/cm$^2$ and $X_0$ is the radiation length, also in g/cm$^2$. The above equation was taken from the PDG journal(\textit{2010} pg 290, 27.14) on Multiple scattering through small angles.

The simulations for investigating the foil materials were run in GEANT4 using the same code for previous electron radiography investigations. A pencil beam of 12 GeV electrons was fired at a slab of material with a detector situated just beyond the object for capturing deflected electrons. Secondary particles were ignored in the detector for these simulations. The geometry is shown in the figure below.

\begin{figure}[H]
\centering
\includegraphics[width=100mm]{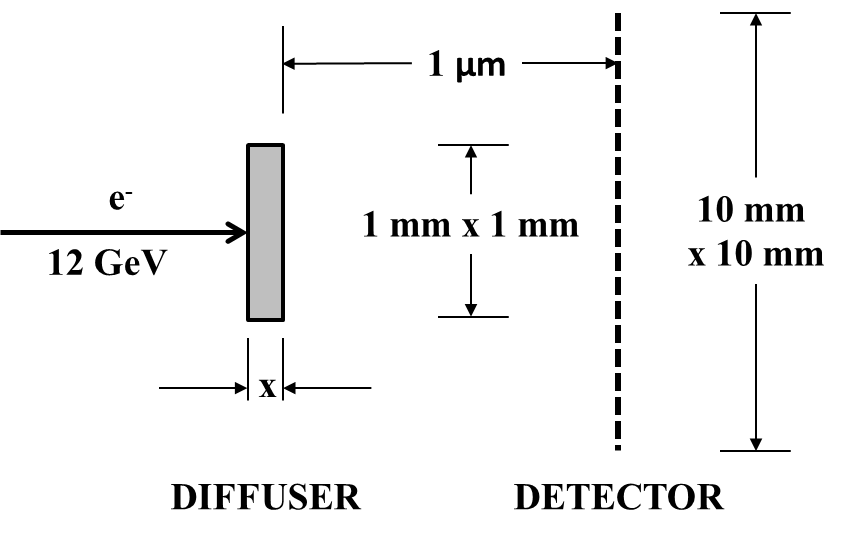}
\caption{The test simulation geometry. Figure is not to scale.}
\end{figure}

Using Equation (1), the initial material thicknesses were used for preliminary simulations. With these initial simulations the angular distribution of the beam through the diffuser was plotted in ROOT to verify the spread was 0.2 mRad. The thickness of the material slab was then incrementally adjusted until the resulting angular spread was 0.2 $\pm$ 0.09 mRad. The calculated and adjusted thicknesses are presented in the table below.

\begin{center}
\begin{tabular}{ |c|c|c|c|c| }
\hline
Material & Z & Melting & Calculated & Adjusted \\ 
 & number & Point (K) & Thickness (m) & Thickness (m)\\
\hline
Graphite & 6 & 3600** & 0.007736 & 0.006900 \\
Silicon & 14 & 1687 & 0.003847 & 0.003100 \\
Titanium & 22 & 1941 & 0.001470 & 0.001150 \\
Iron & 26 & 1811 & 0.000722 & 0.000600 \\
Tantalum & 73 & 3293.15 & 0.000168 & 0.000120 \\
Tungsten & 74 & 3695 & 0.000144 & 0.000110 \\
Platinum & 78 & 4098 & 0.000125 & 0.000090 \\
Uranium & 92 & 1408 & 0.000129 & 0.000095 \\ \hline
 & & **Sublimation & & \\
 & &  Temperature & & \\ \hline
\end{tabular}
\end{center}

\section{Plots and Analysis}

For each diffuser material, a histogram of the energy loss of the electron beam through the material was created. When compared along the same energy range it can be seen that for the lower Z materials there is a much larger energy loss, evident in the RMS values shown in the plots below. The peak of the histogram can be seen to decrease as Z number decreases.

\begin{figure}[H]
\centering
\includegraphics[width=130mm]{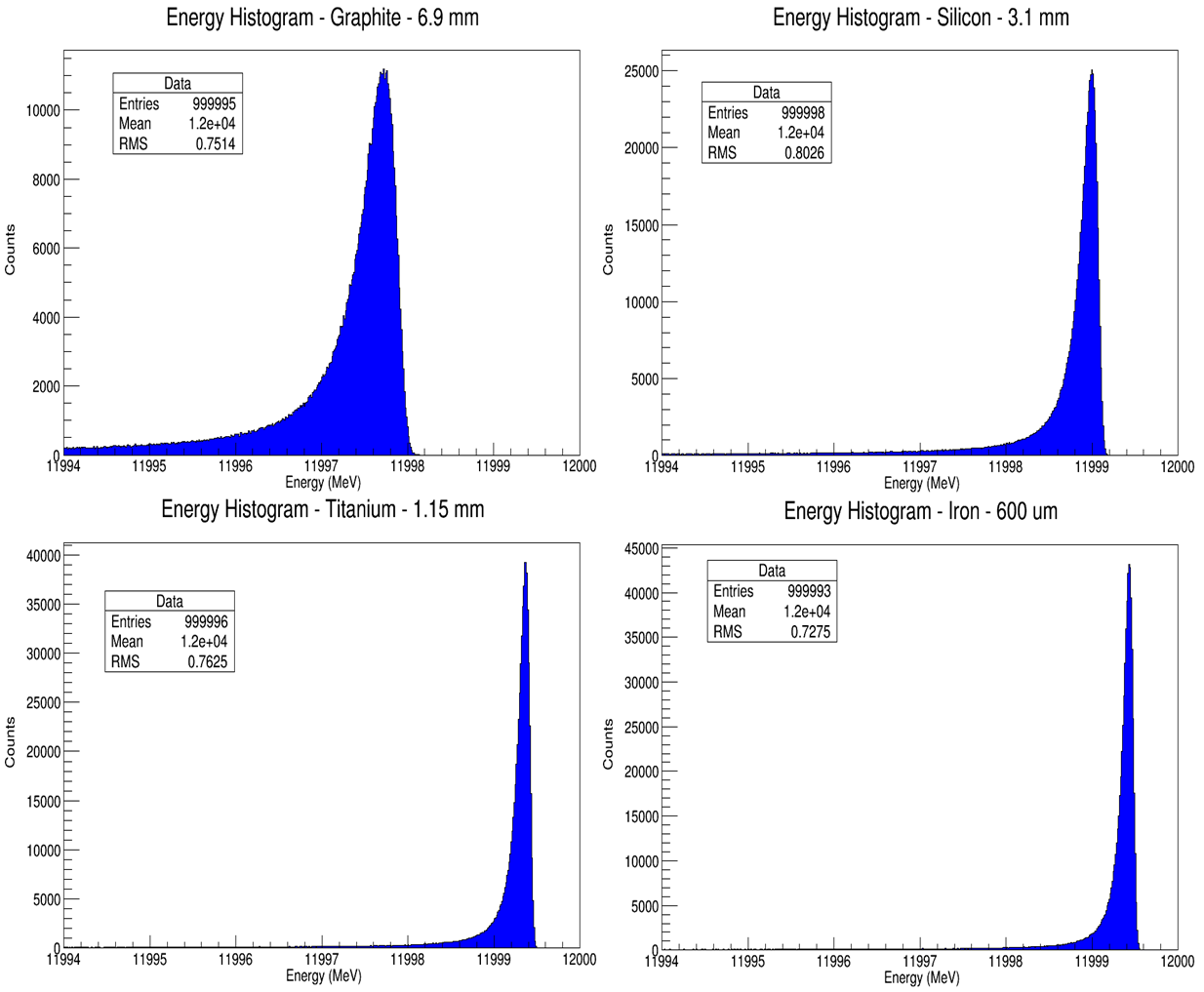}
\caption{Energy loss histograms plotted for the lower Z materials. The energy scale was focused on the range of 11994 MeV to 12000 MeV, with 500 bins.}
\end{figure}

\begin{figure}[H]
\centering
\includegraphics[width=160mm]{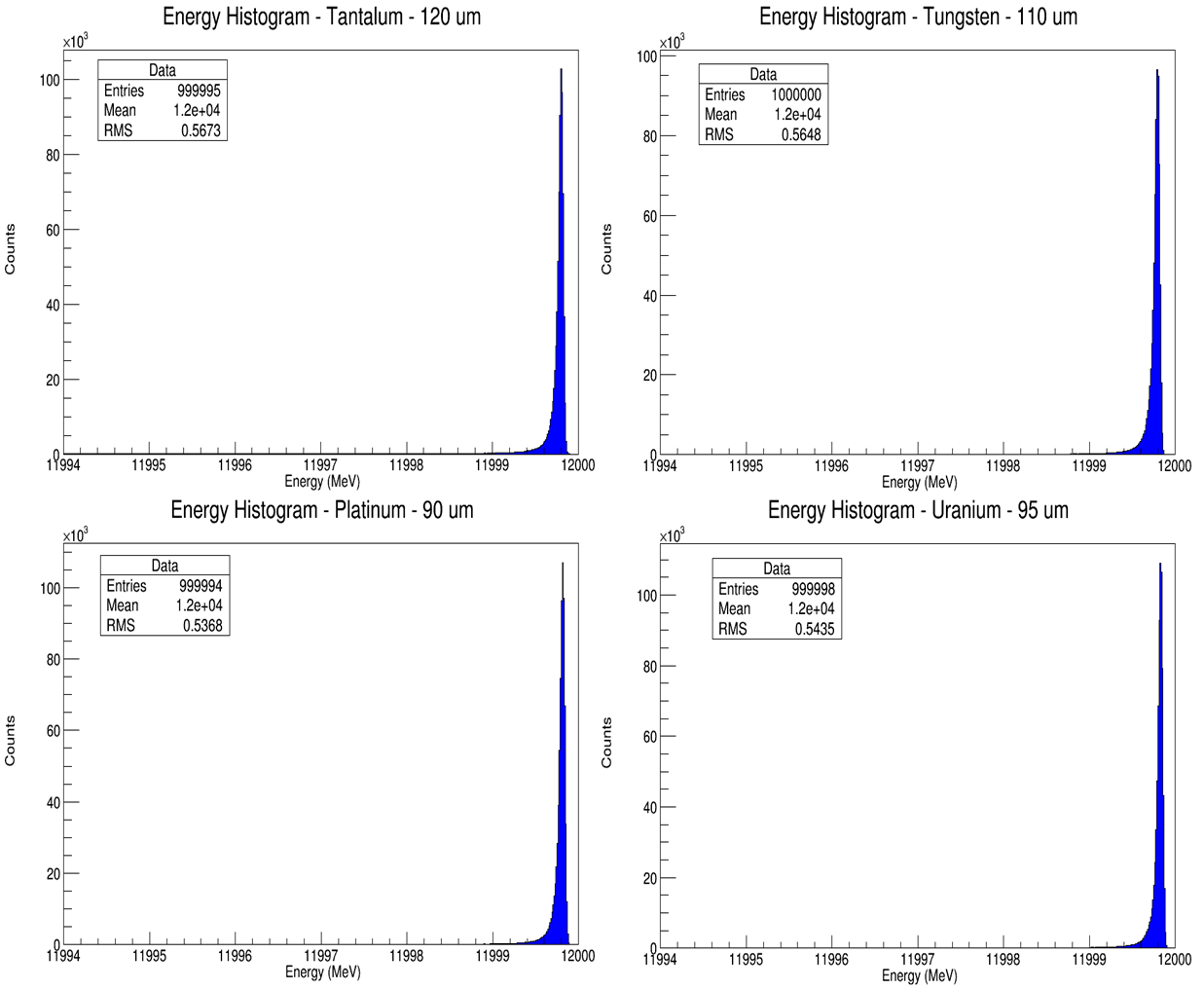}
\caption{Energy loss histograms plotted for the higher Z materials. The energy scale was focused on the range of 11994 MeV to 12000 MeV, with 500 bins.}
\end{figure}

Higher Z materials evidently have less energy loss, and will thus make for better diffusers in electron radiography as the chromatic blur effects will be lessened than with lower Z. To examine a better comparison of the higher Z materials the energy histograms were replotted in a smaller interval to focus on the peak area. From the newer energy plots it is apparent that the peaks and RMS values are close enough that there is no appreciable difference and the choice of diffuser material will depend on other criteria, such as availability of material in foil form, or secondary particle creation. Further investigative studies will be required.  

\begin{figure}[H]
\centering
\includegraphics[width=160mm]{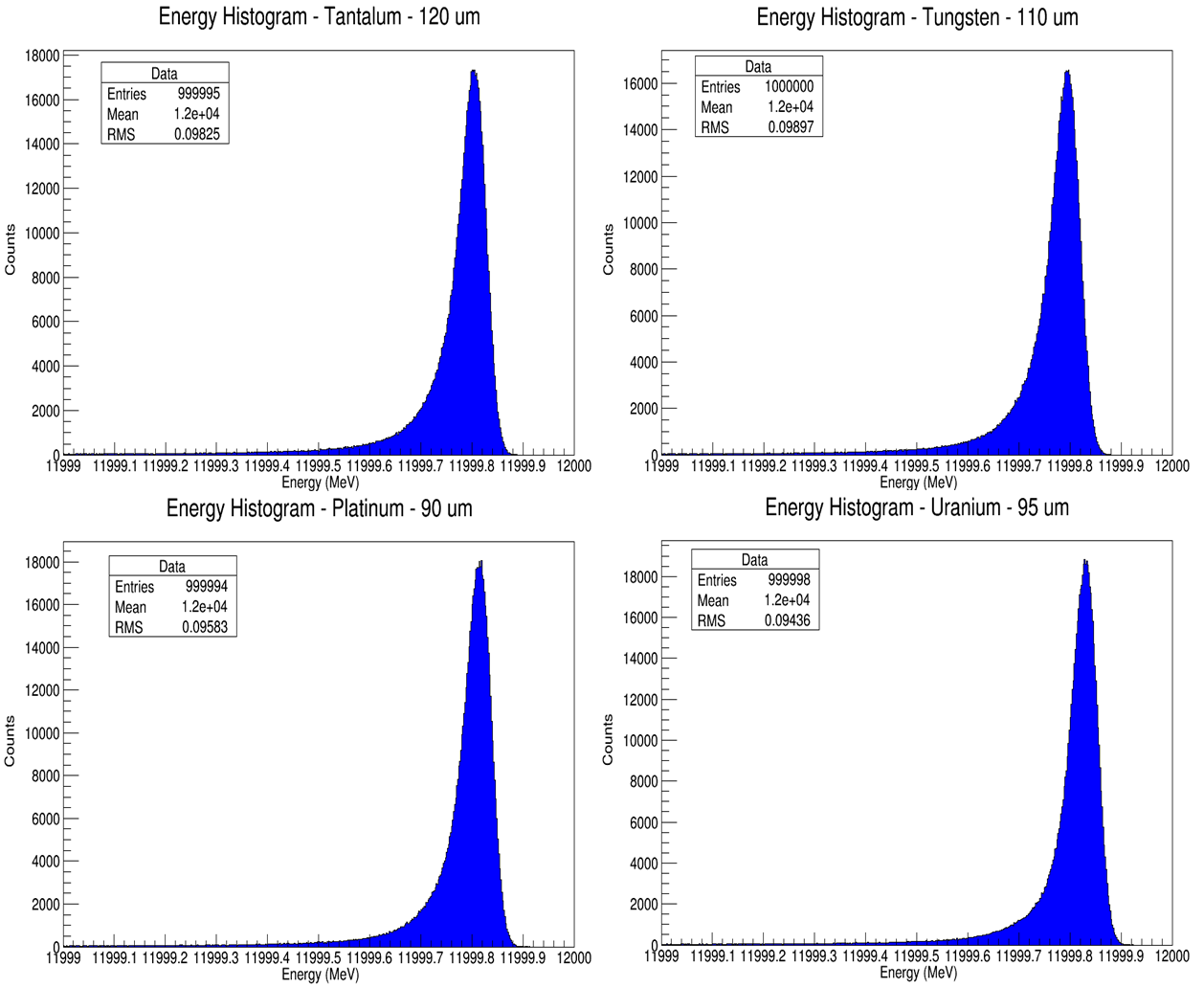}
\caption{Energy loss histograms plotted for the higher Z materials. The energy scale was focused on the range of 11999 MeV to 12000 MeV, with 500 bins.}
\end{figure}

\section{GEANT4 Physics Investigation}

Apart from the refractory nature of the material, the three other processes are important in our investigation. The Multiple Coulomb scattering effect is described by Equation (1) above. 

The deterministic part of the ionization energy loss $\frac{dE}{dx}_\textit{ion.}$ (in units of MeV-cm$^2$/g) has a somewhat more complicated material dependence, including some dependence on the mean ionization potential of the material, but the main factor is trend is that $\frac{dE}{dx}_\textit{ion.}$ increases as $\frac{Z}{A}$ increases. This dependence is illustrated by Fig.  5, which is a plot of the minimum $\frac{dE}{dx}$ for various elements $vs.$ $\frac{Z}{A}$. Random ionization energy straggling, which is added to the deterministic energy loss, giving a Landau distribution for thin objects, also increases as $\frac{Z}{A}$ increases. 

Bremsstrahlung energy loss is small compared to ionization energy loss for thin foils.  In the thick limit, where an electron emits a substantial number of ``hard'' photons, total bremsstrahlung energy loss is proportional to beam energy, i.e. $\frac{dE}{dx} \approx \frac{E_0}{X_0}$, where $E_0$ is the incident energy. However, for typical diffuser foils, we are in the ``thin'' limit, where the probability of a ``hard'' bremsstrahlung event is low (the definition of a hard event is somewhat arbitrary, but it can be taken to be emission of a photon with an energy of 0.1\% of the incident electron energy). 

\begin{figure}[H]
 \centering
 \includegraphics[width=4.0in]{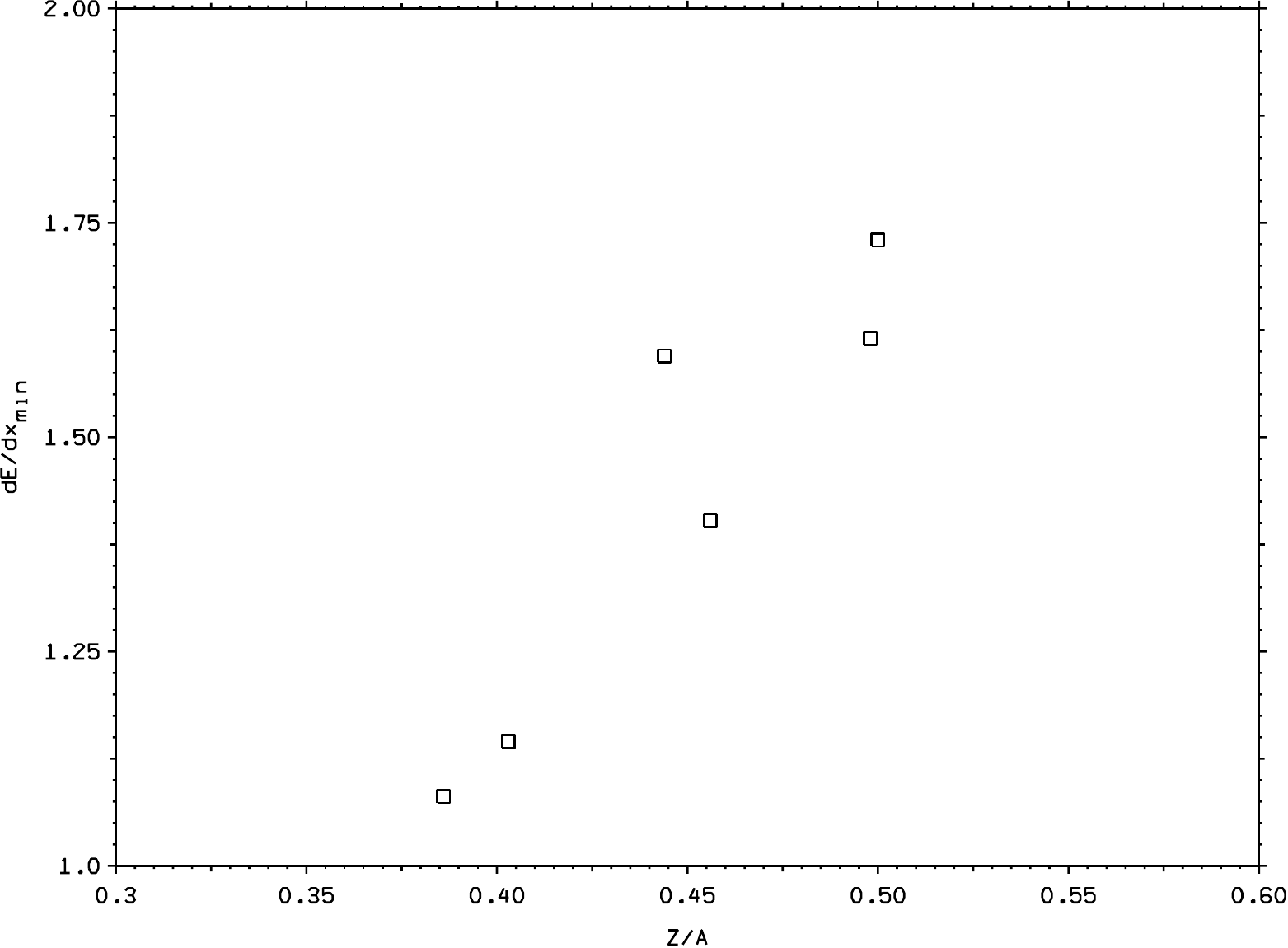}
 \caption{Miniumum ionization $\frac{dE}{dx}$ in MeV-cm$^2$/g vs. $\frac{Z}{A}$ for various materials.  The materials in order of increasing 
 $\frac{Z}{A}$ are U, W, Be, Cu, Al, and C. The outlier is Be. } 
 \label{matchrays}
 \end{figure}

Using the MCS mean-angle formula and ignoring the log factor, we can write for the foil thicknesss $x_{\textit{foil}}$ required to get a certain mean MCS angle $\theta_0$, $x_{\textit{foil}}=C(E)X_0\theta_0^2$, where $C(E)$ is approximately material-independent and contains the dependence on the electron energy. On the other hand, the ionization energy loss distribution for a particular foil thickness $x$ scales roughly as $\frac{Z}{A}$, so the ionization energy loss in a foil of a particular material with a thickness that gives a specified mean scattering angle $\theta_0$  is $\Delta E_{\textit{ion}} \sim \frac{Z}{A} \hspace{0.05in} X_0 \theta_0^2$. Since both $\frac{Z}{A}$ and $X_0$ decrease  with increasing atomic weight, this favors high-Z diffuser materials, provided that their melting temperature is high. 

To investigate the dominant effect in electron deflection within the GEANT4 code a simple scheme was developed. Using the same simulation set-up as the diffuser investigation, a simplified Physics List was written that only included the processes G4eBremsstrahlung and G4eMultipleScattering. Simulations consisted of firing 1 million electrons at a 168 $\mu$m slab of tantalum. Three separate simulations were run with only bremsstrahlung, only multiple Coulomb Scattering, and both processes active. Histograms of the angular spread were plotted and are presented below. It is obvious from the plots that angular deflection is dominated by multiple Coulomb scattering, with electron bremsstrahlung only contributing a small amount to the deflection of the electron as it travels through the tantalum.

\begin{figure}[H]
\centering
\includegraphics[width=150mm]{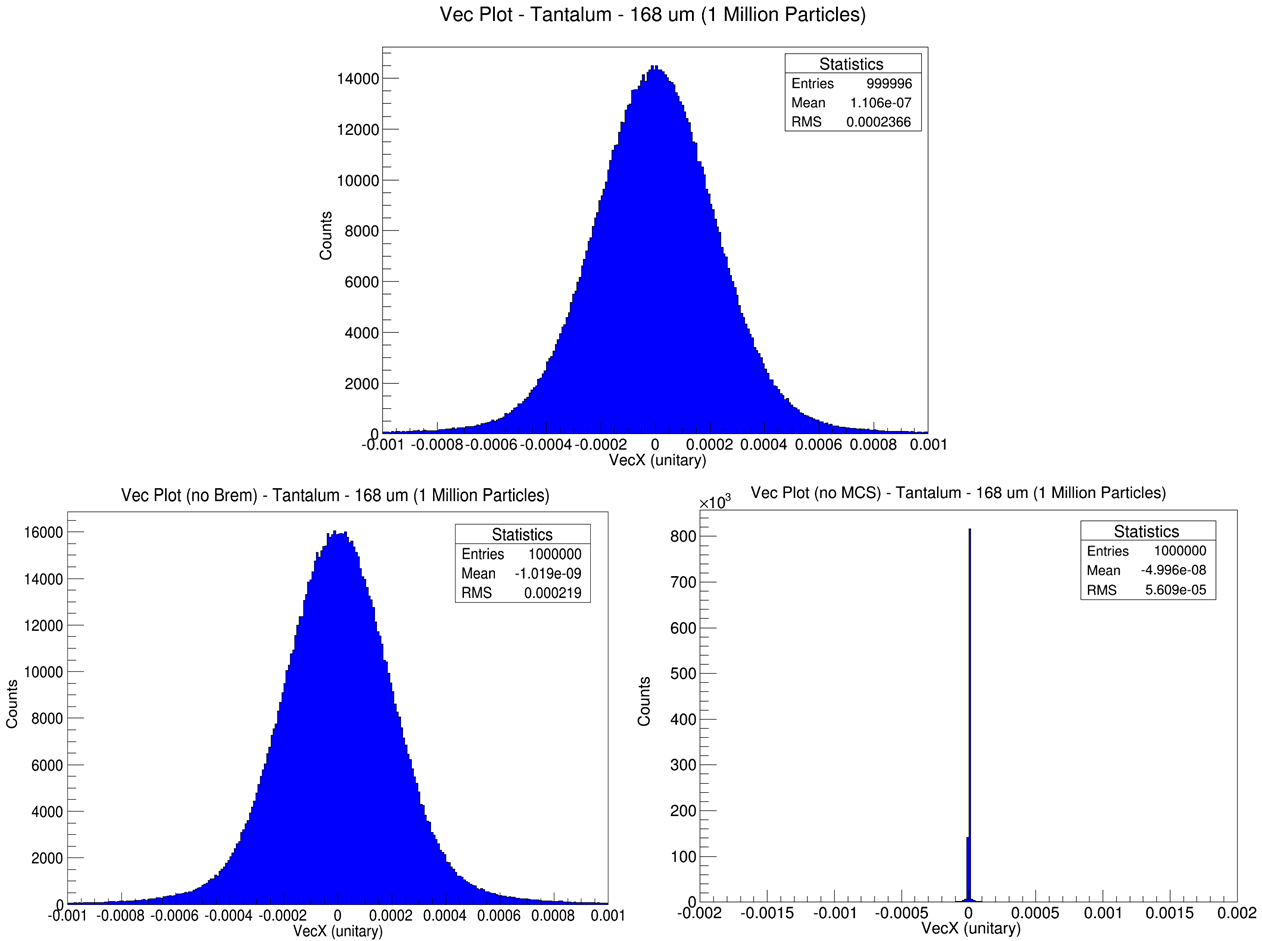}
\caption{Histograms of angular distribution. \textbf{TOP}: Both the G4eBremsstrahlung and G4eMultipleScattering processes were active for this simulation. \textbf{BOTTOM LEFT}: Only the G4eMultipleScattering process was active for this simulation. \textbf{BOTTOM RIGHT}: Only the G4eBremsstrahlung process was active for this simulation.} 
\end{figure}

After investigating the angular spread effects of the physical processes in the GEANT4 code we also wanted to confirm the energy loss effects of those processes. The prior manufactured physics list was modified to include the G4eIonisation process and simulations were run with all three processes, and with only G4eIonisation active. The results, shown in the figure below, confirm that the energy loss of the electrons through the tantalum sample is dominated by the ionization process. 

\begin{figure}[H]
\centering
\includegraphics[width=150mm]{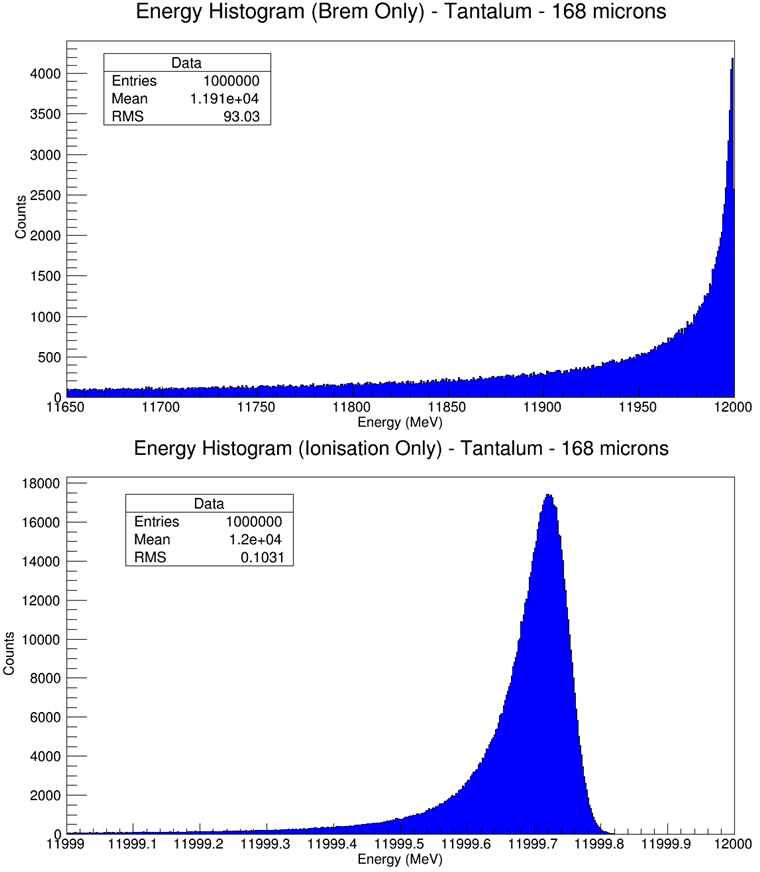}
\caption{Histograms of energy loss. \textbf{TOP}: Only G4eBremsstrahlung active, the majority of electrons(~67.5$\%$) did not lose energy and passed right through the foil. \textbf{BOTTOM}: Only G4eIonisation active. This is the dominant effect on energy loss, as seen when compared to the energy loss diagrams using a full physics list.} 
\end{figure}

We were also curious about the angular dependence of the energy loss from bremsstrahlung in GEANT4. Histograms were created by plotting logarithmic angle versus logarithm of total energy and subtracting the electrons final energy at the detector. The final plot shows there is a correlation between energy loss and angle, so another simulation was run using 12 MeV electrons instead of 12 GeV electrons to see how the correlation would change or if the relation was a static product of a random distribution. Both plots are presented below, and it can be seen that the relation becomes steeper for higher energy particles.

\begin{figure}[H]
\centering
\includegraphics[width=130mm]{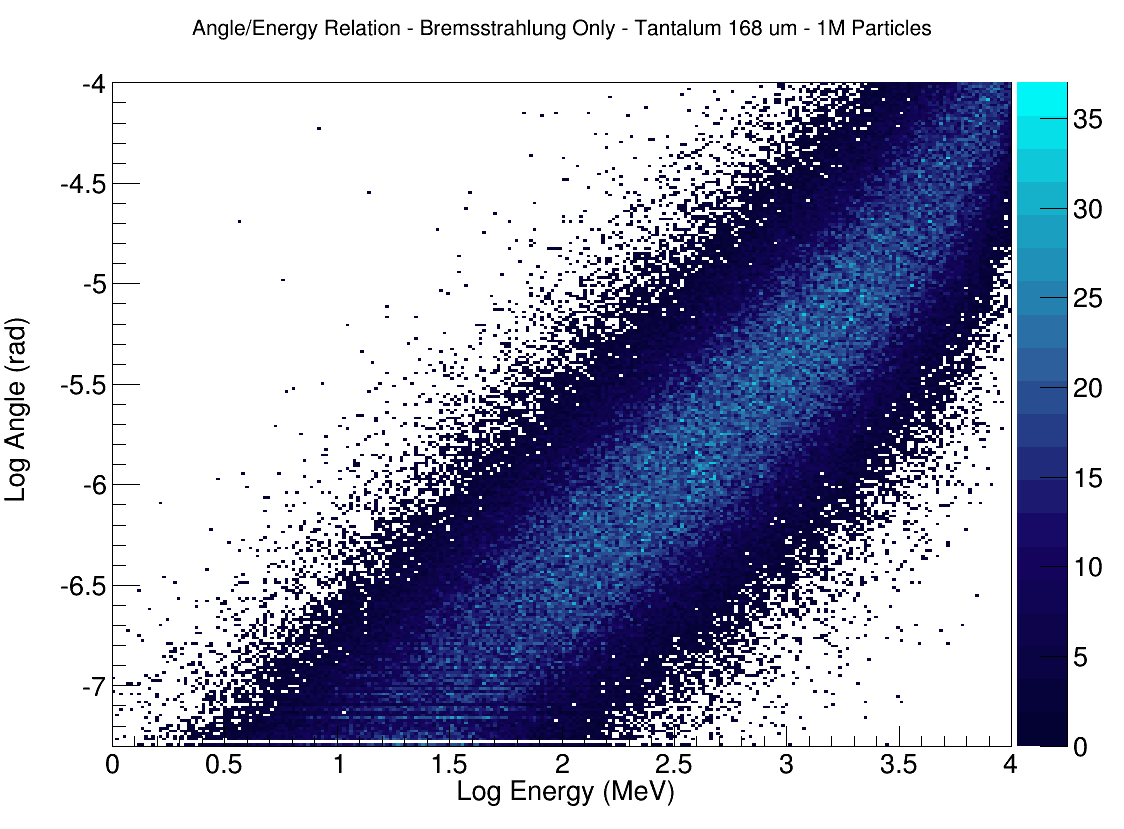}
\caption{Histogram of angle distribution versus energy loss for 12 GeV electron beam incident on 168 $\mu$m tantalum slab.} 
\end{figure}

\begin{figure}[H]
\centering
\includegraphics[width=130mm]{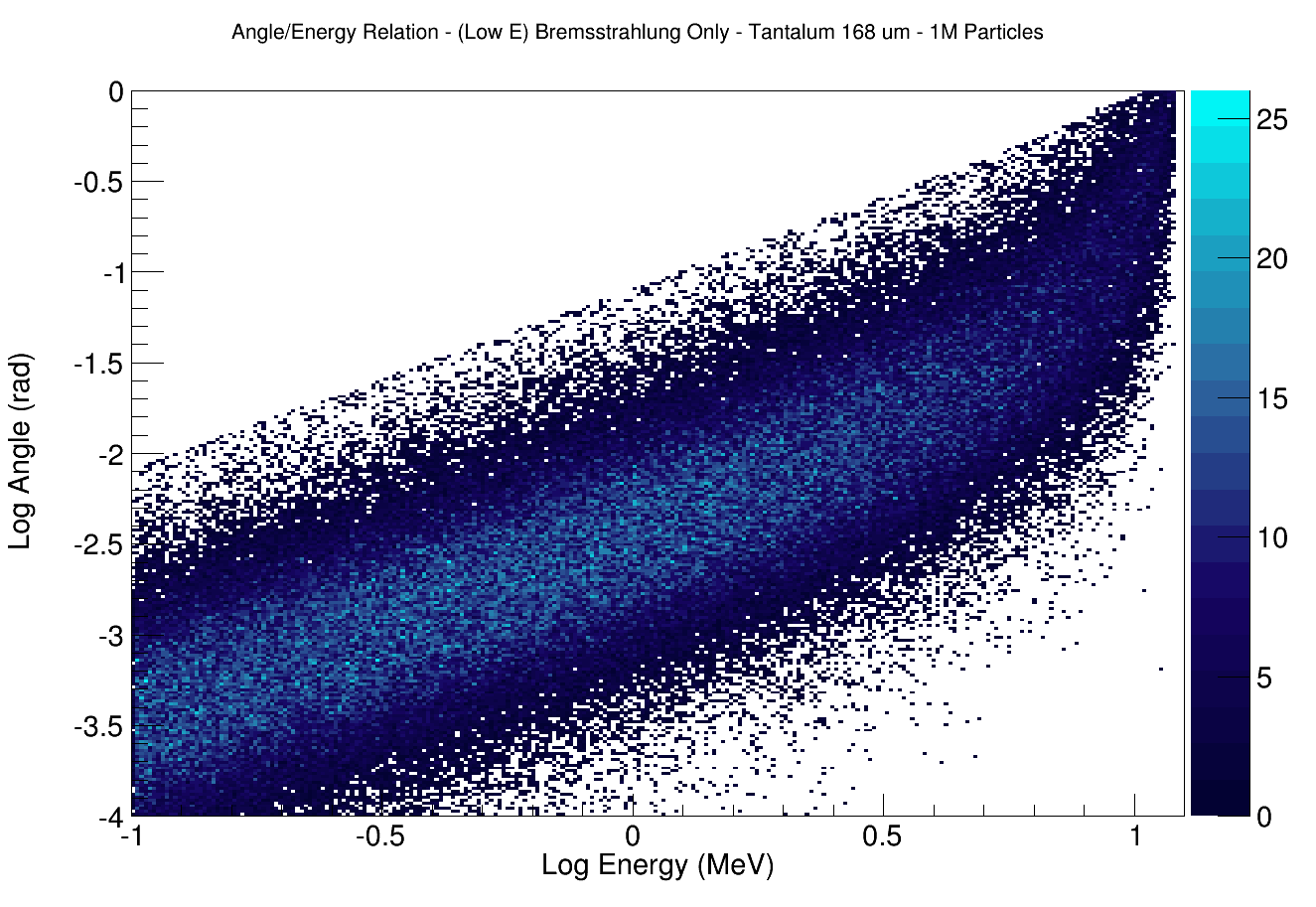}
\caption{Histogram of angle distribution versus energy loss for 12 MeV electron beam incident on 168 $\mu$m tantalum slab.} 
\end{figure}

\end{document}